\documentclass[showpacs,showkeys,preprintnumbers,amsmath,amssymb,twocolumn,pre,floatfix]{revtex4}

\usepackage{amssymb,amsfonts}
\usepackage{amsmath}
\usepackage{graphicx}
\graphicspath{{Figures/}}
\begin{document}

\title{Superradiance Transition in Graphene}

\author{Alexander I. Nesterov}%
   \email{nesterov@cencar.udg.mx}
\affiliation{Departamento de F{\'\i}sica, CUCEI, Universidad de Guadalajara,
Av. Revoluci\'on 1500, Guadalajara, CP 44420, Jalisco, M\'exico}

\author{Ferm\'in Aceves de la Cruz}%
   \email{fermin771009@gmail.com}
\affiliation{Departamento de F{\'\i}sica, CUCEI, Universidad de Guadalajara,
Av. Revoluci\'on 1500, Guadalajara, CP 44420, Jalisco, M\'exico}

\author{Valeriy A. Luchnikov}
 \email{valeriy.luchnikov@uha.fr}
\affiliation{Institut de Science des Matriaux de Mulhouse, UMR 7361 CNRS-UHA, 15 rue Jean Starcky, 68057 Mulhouse-France}

\author{Gennady P.  Berman}
 \email{gpb@lanl.gov}
\affiliation{Theoretical Division, T-4, Los Alamos National Laboratory, and the New
Mexico Consortium,  Los Alamos, NM 87544, USA}

\date{\today}

\begin{abstract}
We study theoretically and numerically the conditions
required for the appearance of a superradiance transition in graphene. The electron properties  of graphene  are described in the single  $p_z$-orbital tight-binding approximation, in which the model is reduced to the effective two-level pseudo-spin $1/2$ system. For each level we introduce the electron transfer rate of escape into a continuum. We demonstrate that, under some conditions, the superradiance experiences the maximal quantum coherent escape to the continuum.
\end{abstract}

\pacs{  42.50.Nn, 72.80.Vp, 03.65.Vf}

\keywords{Superradiance, graphene}

\preprint{LA-UR-15-24090}

\maketitle

\section{Introduction}

The superradiance transition (ST) was first described by Dicke in 1954 \cite{Dicke}.  The ST is normally associated with a significant enhancement of the spontaneous radiation due to quantum coherent effects. 
 Later it was demonstrated that the ST occurs in many quantum optical systems, nuclear systems (heavy nuclei decay), nano- and bio-systems \cite{VZA,VZ1,VZ2,Zel1,VZ3,Ber3}. The ST usually occurs when the discrete (intrinsic) states of the system interact with the continuum spectra (sinks). Then, an adequate approach for describing the eigenstates and the dynamics of the system can be based on an effective non-Hermitian Hamiltonian for intrinsic states \cite{VZA,VZ1,VZ2,Zel1,VZ3,Ber3}. In this case, the eigenenergies of the non-Hermitian Hamiltonian  become complex. Recently, a  stady-state superradiant laser with less than one intracavity photon was demonstrated with rubidium-87 atomic dipoles \cite{SL}.  Large enhancement of F\"orster resonance energy transfer on graphene platforms was discussed in \cite {Ag}.  In \cite{Ryzhii}, a superradiant plasmonic lasing with a giant gain at the palsmon modes in graphene was theoretically analyzed in a wide THz frequency range.

Qualitatively, the ST occurs when the resonances begin to overlap -- the spacing between the resonances becomes of the order of the sum of half-widths of these resonances. With further overlapping of resonances, segregation of the eigenenergies takes place, depending on their decay widths. Namely, the wide superradiant eigenstates provide rapid and coherent decay of the initially populated state in the continuum. The subradient eigenstates, with narrow decay widths, survive for a relatively long time. Note, that not only the overlapping of resonances and the segregation of the eigenenergies are important for the occurrence of the superradiance, but the initial population of the system is also important. All these effects are described in details in \cite{VZA,VZ1,VZ2,Zel1,VZ3,Ber3,BNLS}. 

 In this letter, we determine theoretically and numerically the conditions at which the ST occurs in a single-layer graphene material. We demonstrate how the ST in this system is related to both the occurrence of an exceptional points (EP), when complex eigenvalues coincide, and to the overlapping of two resonances. We also show that, under some conditions on the parameters and initial populations, the maximal efficiency of the electron transfer (ET) into the sink is related to the ST. We compare the ST criterion based on the overlapping of resonances and on the occurrence of the EP. We 

\section{Description of the model}

\begin{figure}
\scalebox{0.2}{\includegraphics{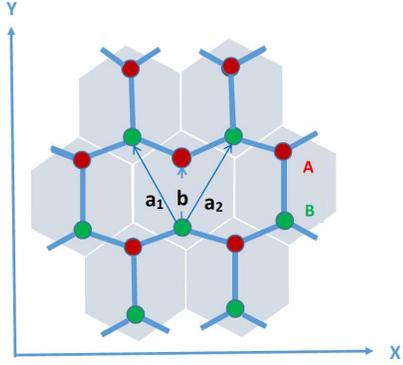}}	
\caption{(Color online) Graphene two-dimensional lattice, with two atoms, $  A$ (red) and $ B $ (green), in a unit cell. Primitive unit vectors: $\mathbf a_{1,2}=(a/2)(\sqrt{3},3)$, $ {\mathbf b}=(0,a)$, $a\approx 1.42 \, \rm \AA$.}
\label{TL}
\end{figure}

The structure of a single atomic layer graphene can be described by the 
honeycomb lattice which consists of two triangular Bravais sublattices, 
represented in Fig. \ref{TL} by nonequivalent A (red)  and B (green) carbon 
atoms, which create a unit cell \cite{Wall,Alev}. Both sublattices have the 
periodic structures, and are shifted by a vector, $ {\mathbf b}=(0,a)$, 
$a\approx 1.42 \, \rm \AA$, which connects A and B atom in the unit cell. 

In a carbon atom, six electrons occupy the $1s^2$, $2s^2$, and $2p$ orbitals.  From them, four valence electrons are responsible for structural and electronic properties. One of the valence electrons of each A and B atoms occupies the $p_z$ orbital, which is orthogonal the graphene plane. The hybridization of these $p_z$ orbitals provides the formation of the $\pi$-bands in graphene. Then, the electron properties of graphene can be described within a single $p_z$ orbital tight-binding (TB) model \cite{Alev}. Using the TB approximation, one can show that the effective single-electron Hermitian Hamiltonian is reduced to the two-level pseudo-spin one-half system. The projections of pseudo-spin are associated with two sublattices. 

\begin{figure}
\scalebox{0.25}{\includegraphics{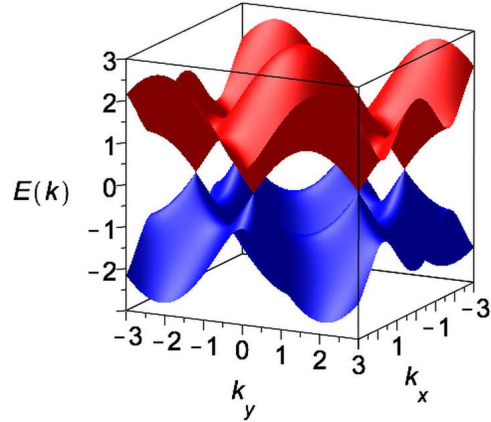}}	
\caption{(Color online) The band structure of graphene. The valence and conduction bands touch each other in six DP points, where  $E(\mathbf k) = 0$.}
\label{BS}
\end{figure}

In graphene, the dispersion relation, $E(\mathbf k)$ (where, $\mathbf k$, is the wave vector), has some specific properties: the Fermi level corresponds to $E(\mathbf k)=0$, and the valence and conduction bands touch each other in the first Brillouin zone at six points. Each of these points provides a ``conical intersection" known also as the ``diabolical point'' (DP) \cite{B0,BW}.  (See Fig. \ref{BS}.)

According to \cite{Alev}, in the $\mathbf k$-representation the dynamics of the electron, in the vicinity of the DP, can be described by the following Hamiltonian:
\begin{eqnarray}\label{GH}
{\mathcal H}_0 = \left(\begin{array}{cc}
0&\hbar v_F(q_x- iq_y) \\
\hbar  v_F( q_x+ iq_y) & 0
\end{array}
\right ),
\end{eqnarray}
where, $ v_F = 3|t|a/2\hbar$, is the Fermi velocity. Here $t \approx -2.8 \,\rm eV$, is the hopping integral for nearest neighbor atoms, A and B, with coordinates, $\mathbf R_A$ and $\mathbf  R_B$.  We use the following notations: the state, $|1\rangle=\big ({\scriptsize\begin{array}{c}
   1\\
   0
   \end{array}}\big )$, corresponds to the population of the sublattice A, and the state, $|2\rangle=\big ({\scriptsize\begin{array}{c}
     0\\
     1
     \end{array}}\big )$, corresponds to the population of the sublattice B. The Hamiltonian, ${\mathcal H}_0$, has the eigenvalues, $E_{\pm} =\pm \hbar v_F|\mathbf q|$. The corresponding eigenstates are:  $|+\rangle=(1/\sqrt{2})(e^{-i\varphi/2}|1\rangle + e^{ i\varphi/2}|2\rangle)$ (conduction band), and  $|-\rangle=(1/\sqrt{2})( e^{-i\varphi/2}|1\rangle - e^{i\varphi/2} |2\rangle)$ (valence band), where $\varphi= \arg(q_x+ iq_y)$.

Suppose that each discrete state (related to the sublattice, A and B) is
coupled to a continuum -- continuous energy band, which could originate due to impurities or other mechanisms. Suppose, that we allow an electron to tunnel to the continuum from the sublattices, A and B, with the ET rates, $\Gamma_1$  and  $\Gamma_2$, correspondingly. Then, the quantum dynamics of the ET can be described by the following effective non-Hermitian Hamiltonian, $ \tilde{\mathcal H}= {\mathcal H}- i \mathcal W$, where, $ {\mathcal H} $,
is the dressed Hamiltonian, $ {\mathcal H}_0 $, and
\begin{eqnarray}
\label{Gamma}
{\mathcal W} = \frac{1}{2} \left(\begin{array}{cc}
\Gamma_1&0\\
0 & \Gamma_2
\end{array}
\right). 
\end{eqnarray}
We find,
\begin{eqnarray}\label{Ham1}
\tilde  {\mathcal H} = \frac{\lambda_0}{2}\left(\begin{array}{cc}
1&0\\
0 & 1
\end{array}
\right ) +  \frac{1}{2}\left(\begin{array}{cc}
\varepsilon - i \Gamma &V^\ast\\
V &  -\varepsilon + i \Gamma
\end{array}
\right),
\end{eqnarray}
where, $\lambda_0= \varepsilon_0 - i\Gamma_0$,   $\varepsilon_0 = \varepsilon_1 + \varepsilon_2$, $\Gamma_0 =(\Gamma_1 + \Gamma_2 )/2$, $V= 2\hbar v_F(q_x + iq_y)$, $\varepsilon = \varepsilon_1 - \varepsilon_2$, and  $\Gamma =(\Gamma_1 - \Gamma_2 )/2$. Here, $\varepsilon_n$, is the renormalized energy of the state, $|n\rangle$, which usually occurs in the non-Hermitian Hamiltonian approach \cite{VZA,VZ1,VZ2,Zel1,VZ3}.   Note, that for pure graphene at the DP, the ``electron masses", $\varepsilon_{1,2}=0$, for both sublattices. However, due to the finite bandwidths, associated with the sinks, not only the ET rates, $\Gamma_{1,2}$, appear, but also the small ``effective electron masses'', $\varepsilon_{1,2}$, may occur.  
 
Below, in numerical simulations, we choose $\hbar=1$. All energy-dimensional parameters are measured in $\rm ps^{-1}\approx 0.66~\rm meV$. Time is measured in $\rm ps$.  For example, if we choose, $|{\mathbf q}|=10^4~\rm cm^{-1}$ and $v_F=10^8~\rm {cm/sec}$, we have: $|V|=2~\rm ps^{-1}\approx 1.32\rm meV$.

\section{Superradiance transition}

The bi-orthogonal eigenstates of the effective non-Hermitian Hamiltonian, $\tilde{\mathcal H}$, provide a convenient basis in which the  eigenvalue problem can be formulated and resolved. The solution of the eigenvalue problem, $ \tilde{\mathcal H}|u\rangle =\tilde E|u\rangle$ and $ \langle \tilde u| \tilde{\mathcal H}= \tilde E\langle \tilde
u|$ (where, $|u\rangle$ and $\langle \tilde u|$, are the right and left eigenvectors
respectively), is given by,
\begin{equation}
\tilde E_{1,2} = \lambda_0/2 \pm \Omega/2, \label{EigL}
\end{equation} 
where, $\Omega= \sqrt{|V|^2 +(\varepsilon - i\Gamma )^2}$. The right and the left eigenstates form the bi-orthonormal basis with the properties:
$\langle \tilde u_{1,2}|u_{2,1}\rangle = 0 $, $ \langle \tilde
u_{1,2}|u_{1,2}\rangle = 1 $. 

\begin{figure}
 \scalebox{0.25}{\includegraphics{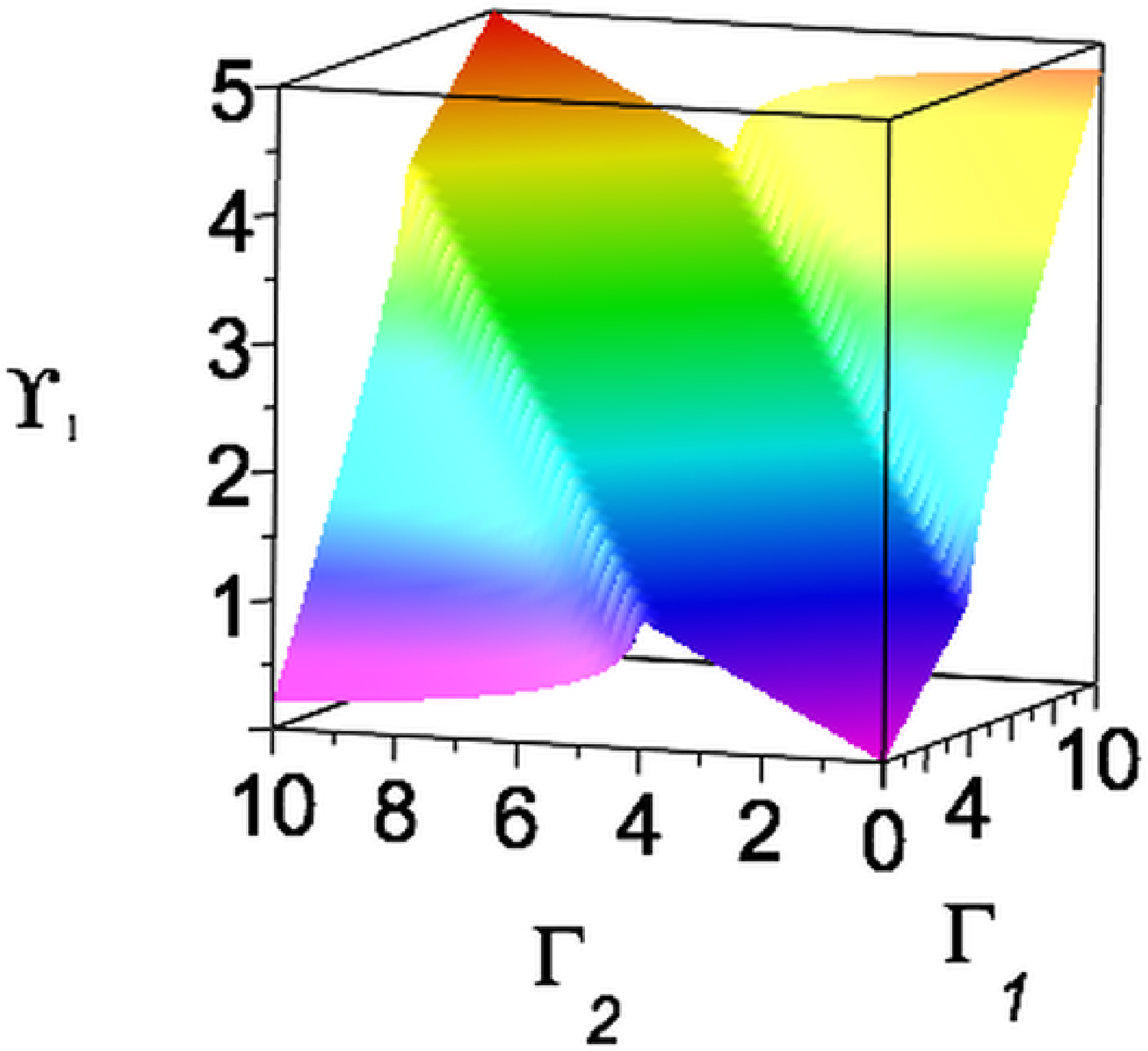}}	
 (a)
 \scalebox{0.2}{\includegraphics{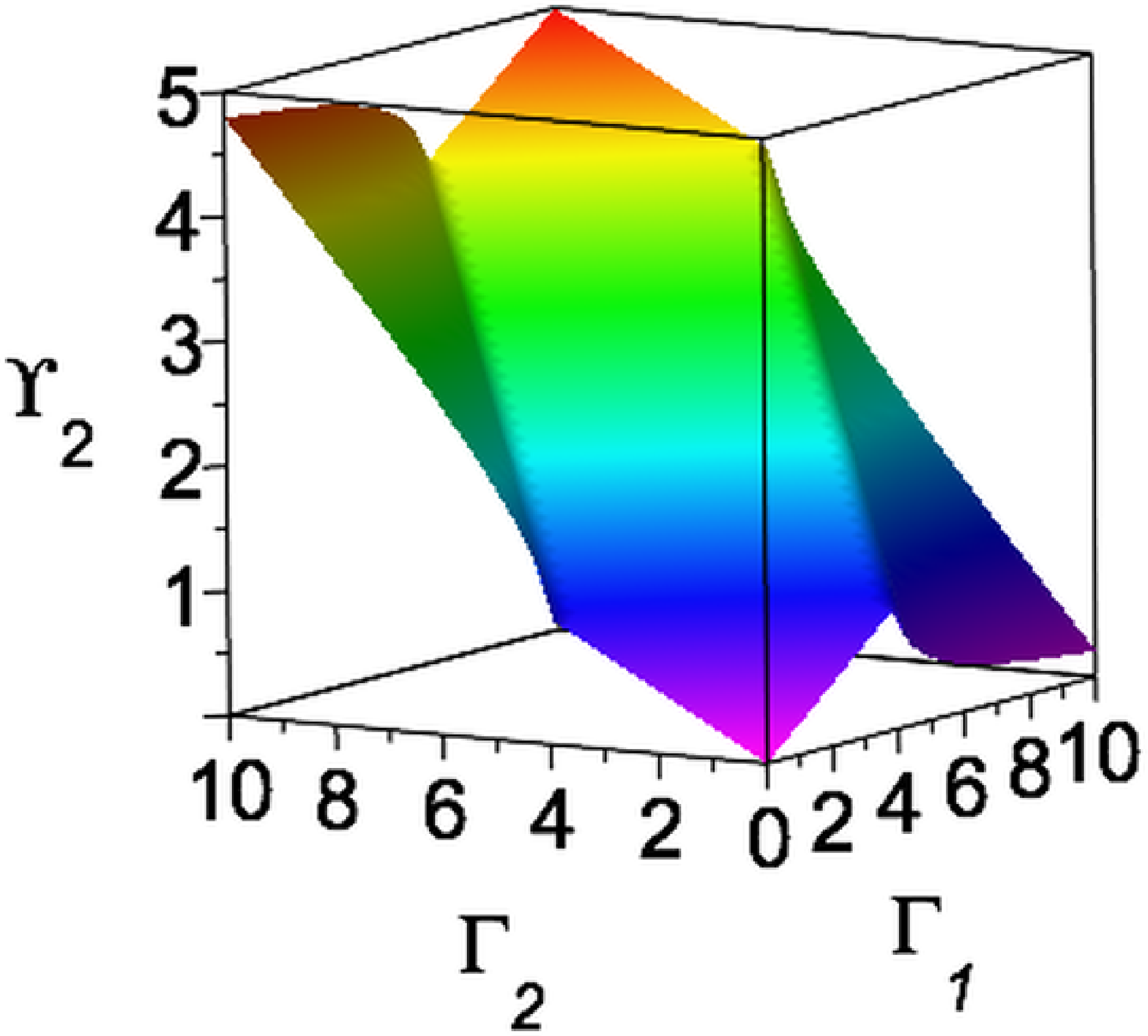}}	
 (b)	
 \begin{center}
 \scalebox{0.25}{\includegraphics{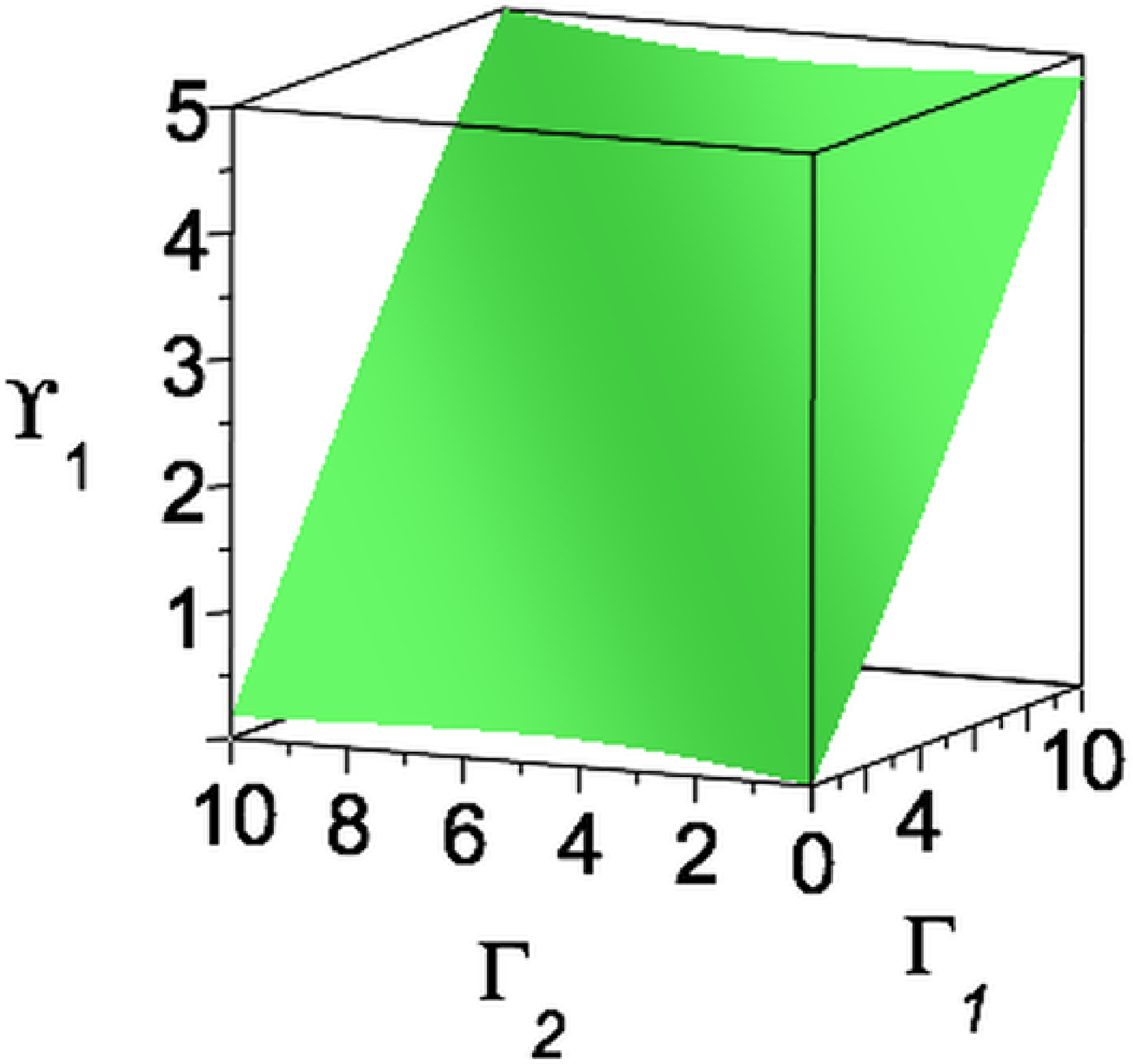}}	
 (c)
 \scalebox{0.25 }{\includegraphics{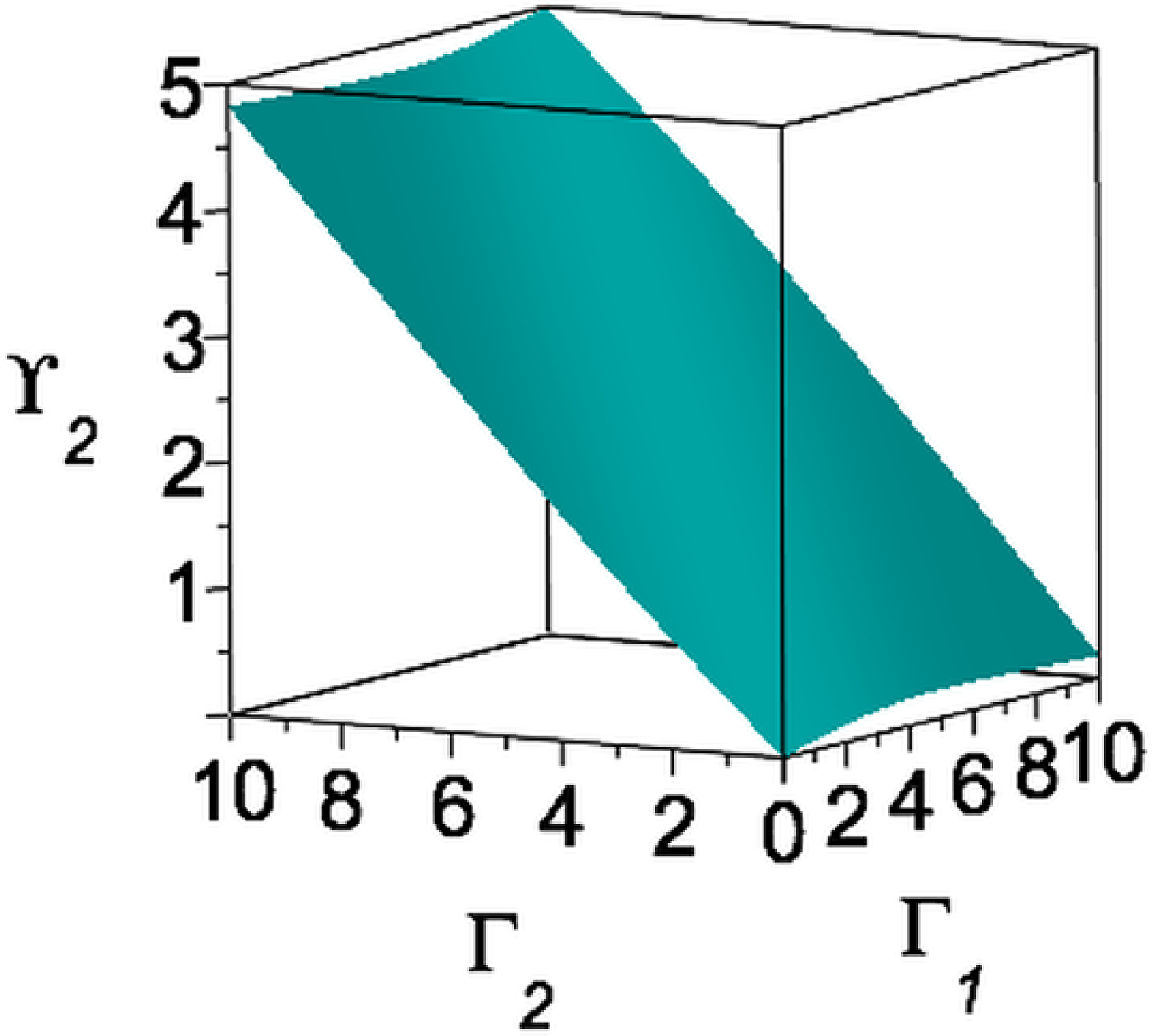}}	
 (d)
 \end{center}
 \caption{(Color online)  Decay widths, $\Upsilon_\alpha$ ($\alpha=1,2$), as the functions of the rates, $\Gamma_1$ and $\Gamma_2$ ($|V|=2$): (a,b) $\varepsilon=0$; (c,d) $\varepsilon=2$. 
 \label{D1}}
 \end{figure}

Further, it is convenient to set, $\tilde E_{\alpha} = {\mathcal E}_\alpha -i\Upsilon_{\alpha}$ ($\alpha=1,2$), where, ${\mathcal E}_\alpha= \Re \tilde E_\alpha$, and $\Upsilon_{\alpha}= -\Im \tilde E_\alpha$ is the value of the complex energy (or the half-width of the resonance, $\alpha$ \cite{VZ1,VZ2,Zel1,VZ3}). In Fig. \ref{D1}, the values, $\Upsilon_\alpha$, as the functions of $\Gamma_1$ and $\Gamma_2$, are presented.

For $\Gamma_{1} =\Gamma_{2}$, one has the  standard energy levels repulsion, $\Delta \mathcal E =\Omega_0 $, where  $\Omega_0= \sqrt{\varepsilon^2 +|V|^2  }$, and $\Delta \mathcal E = \mathcal E_1 - \mathcal E_2$ denotes the energy level spacing.  When $|\Gamma| \ll \Omega_0$,  we obtain,
\begin{eqnarray}
&&\Delta \mathcal E \approx \Omega_0 -  \frac{\Gamma^2}{2\Omega_0}, \\
&&\Upsilon_{1} \approx \frac{\Gamma_0}{2} + \frac{\varepsilon}{\Omega_0}\Gamma, \quad
\Upsilon_{2} \approx \frac{\Gamma_0}{2}  - \frac{\varepsilon}{\Omega_0}\Gamma.
\end{eqnarray}
From here it follows that, while the decay widths separate as $\Gamma$ grows (a segregation effect \cite{Zel1}), the real  energies, ${\mathcal E}_1$ and ${\mathcal E}_2$, are attracted to each other.
 
In the opposite limit, $|\Gamma| \gg \Omega_0$, we find,
\begin{eqnarray}
&&\Delta \mathcal E \approx \varepsilon + \frac{\varepsilon\Omega^2_0}{2\Gamma^2}, \\
&&\Upsilon_{1} \approx \frac{\Gamma_1}{2} - \frac{\Omega^2_0}{2\Gamma}, \quad
\Upsilon_{2} \approx \frac{\Gamma_2}{2}  +\frac{\Omega^2_0}{2\Gamma}.
\end{eqnarray}
One can see that if $\Gamma_1\gg \Gamma_2$, the eigenstate, $|u_1\rangle$, corresponding to $\tilde E_1$, is the superradiant state, and the eigenstate $|u_2\rangle$, corresponding to $\tilde E_2$, is a  long-living subradient state.  In the opposite case, $\Gamma_2\gg \Gamma_1$, the eigenstate, $|u_1\rangle$, becomes long-living subradient state,  and the eigenstate, $|u_2\rangle$, is a rapidly decaying superradiant state.

For two-level system, governed by the effective non-Hermitian Hamiltonian, $\tilde{\mathcal H}$, the qualitative criterion of the overlapping of two resonances can be written as follows: $\Gamma_0/\Delta \mathcal E \approx 1$, which,  after some algebra, reduces to the equation, 
\begin{eqnarray}
\Gamma_0^4 + \Gamma_0^2 \Gamma^2 -\Omega_0^2  \Gamma_0^2 - \varepsilon^2  \Gamma^2=0.
\end{eqnarray}
When $\Gamma_2=0$ (or $\Gamma_1 =0$), the solution of this equation takes a simple form,
$\Gamma_{1,2}^\ast = \sqrt{2(\Omega_0^2 + \varepsilon^2)}$.

\begin{figure}[ht]
\scalebox{0.175}{\includegraphics{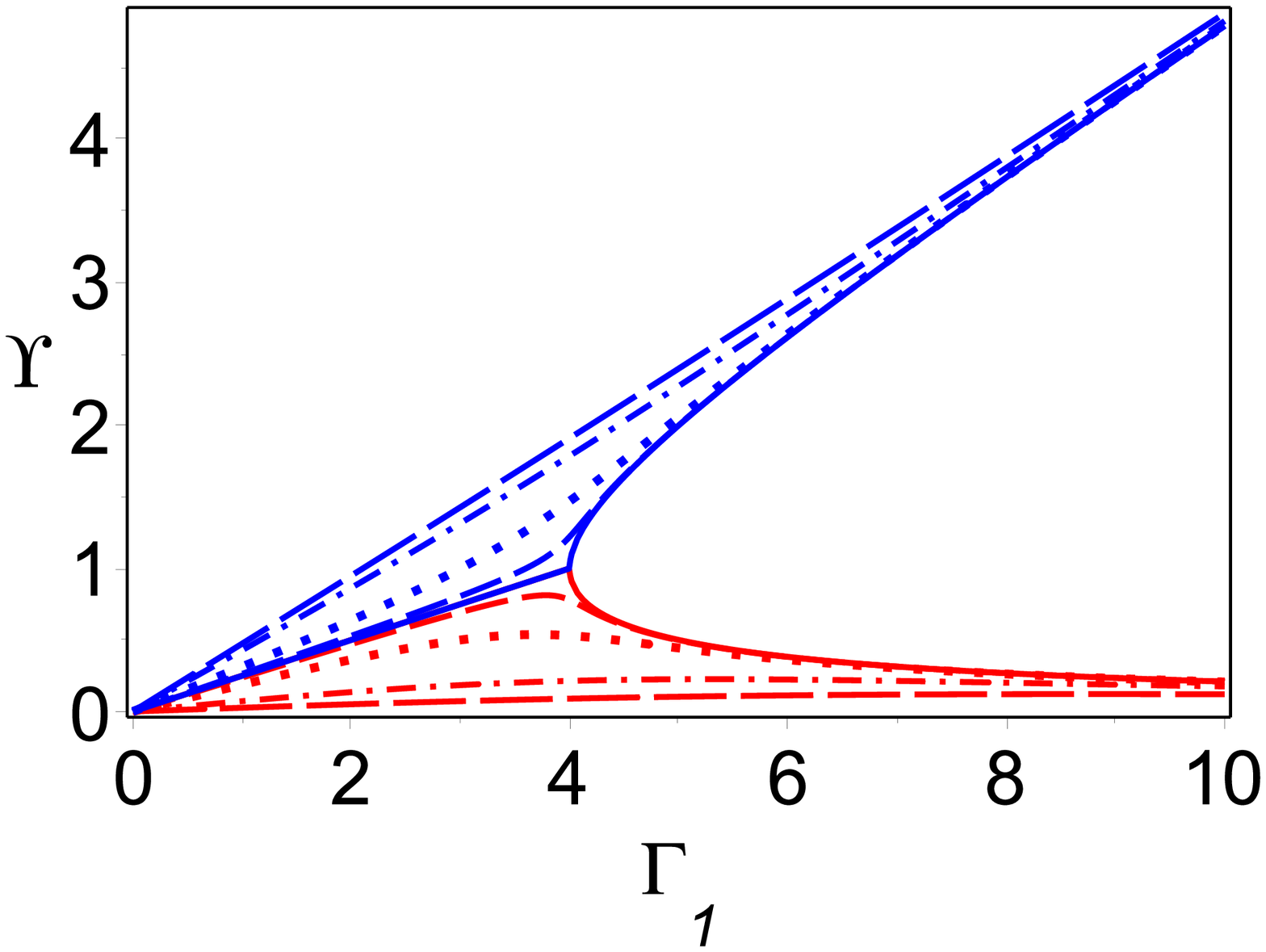}}	
\scalebox{0.18}{\includegraphics{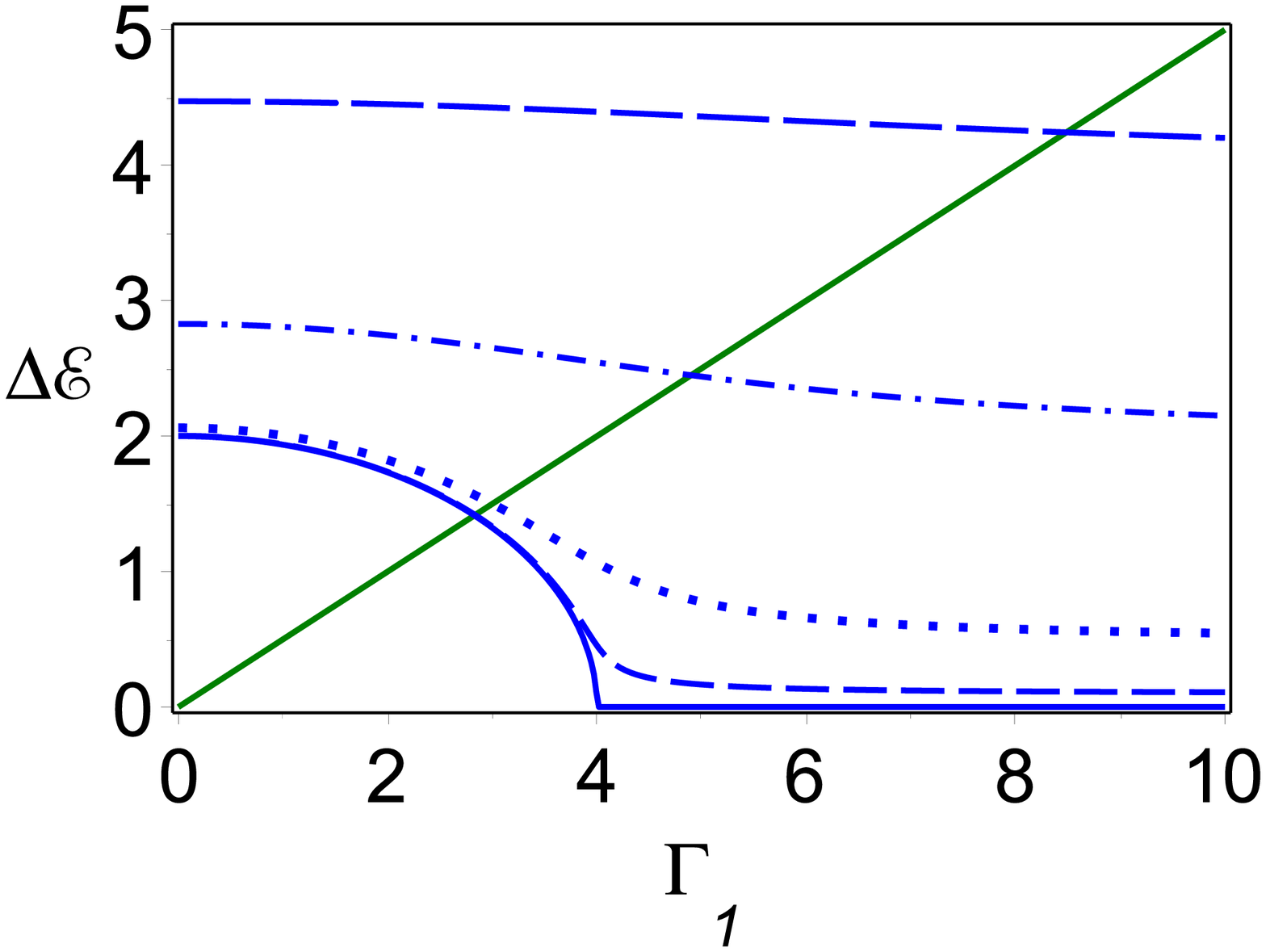}}	
\caption{(Color online) Left panel: $\Upsilon_\alpha$ ($\alpha=1,2$) as the functions of the rate, $\Gamma_1$; $\Upsilon_1$ (blue) and $\Upsilon_2$ (red).  Right panel: Dependence of the spacing, $\Delta \mathcal E$ (blue), on $\Gamma_1$. Green line represents  the sum, $\Upsilon =\Upsilon_1 +\Upsilon_2$, vs $\Gamma_1$.  Parameters:  $\Gamma_2=0$, $|V|=2$, $\varepsilon=0$ (solid curves); $\varepsilon=0.1$,  (dashed curves); $\varepsilon=0.5$ (dotted curves); $\varepsilon=2$ (dot-dashed curves); $\varepsilon=4$ (long-dashed curves).
\label{D1a}}
\end{figure}

In Fig. \ref{D1a}, the half-widths of the resonances, $\Upsilon_\alpha$, and the spacing, $\Delta \mathcal E$, vs $\Gamma_1$ are depicted, for the choice of the ET rate, $\Gamma_2 =0$. The ST occurs at the point, where the smallest half-width, $\Upsilon_2$, reaches its maximum value (Fig.  \ref{D1a}a). One can see that the qualitative criterion of the overlapping of resonances is good enough for $\varepsilon \gtrsim 2$. Indeed, as Fig. \ref{D1a} demonstrates, for $\varepsilon= 2$ (and $|V| =2$), the ST occurs at the point $\Gamma_1 \approx 5.3$, while the criterion of the overlapping of resonances yields, $\Gamma_{1}^\ast\approx 4.9$. The error in the estimate of the ST is, $\approx 7.5 \,\%$.

In the opposite case, of the small values of $\varepsilon\ll1$, as in graphene, the ST corresponds to the EP ($\varepsilon=0$), and the criterion of the overlapping of resonances cannot be used.

\subsection{Superradiance transition in the vicinity of the exceptional point}

\begin{figure}
\scalebox{0.14}{\includegraphics{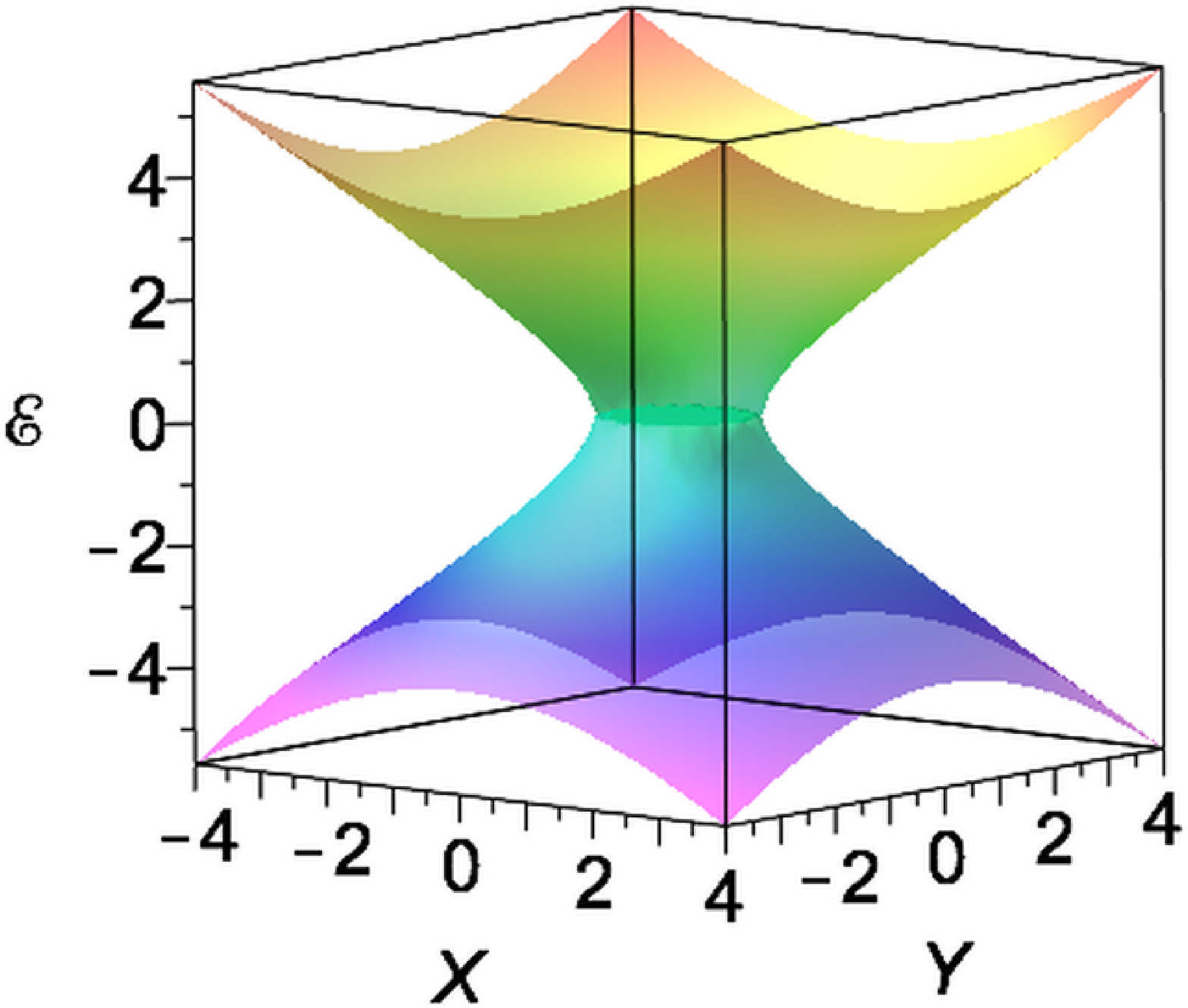}}	
\scalebox{0.14}{\includegraphics{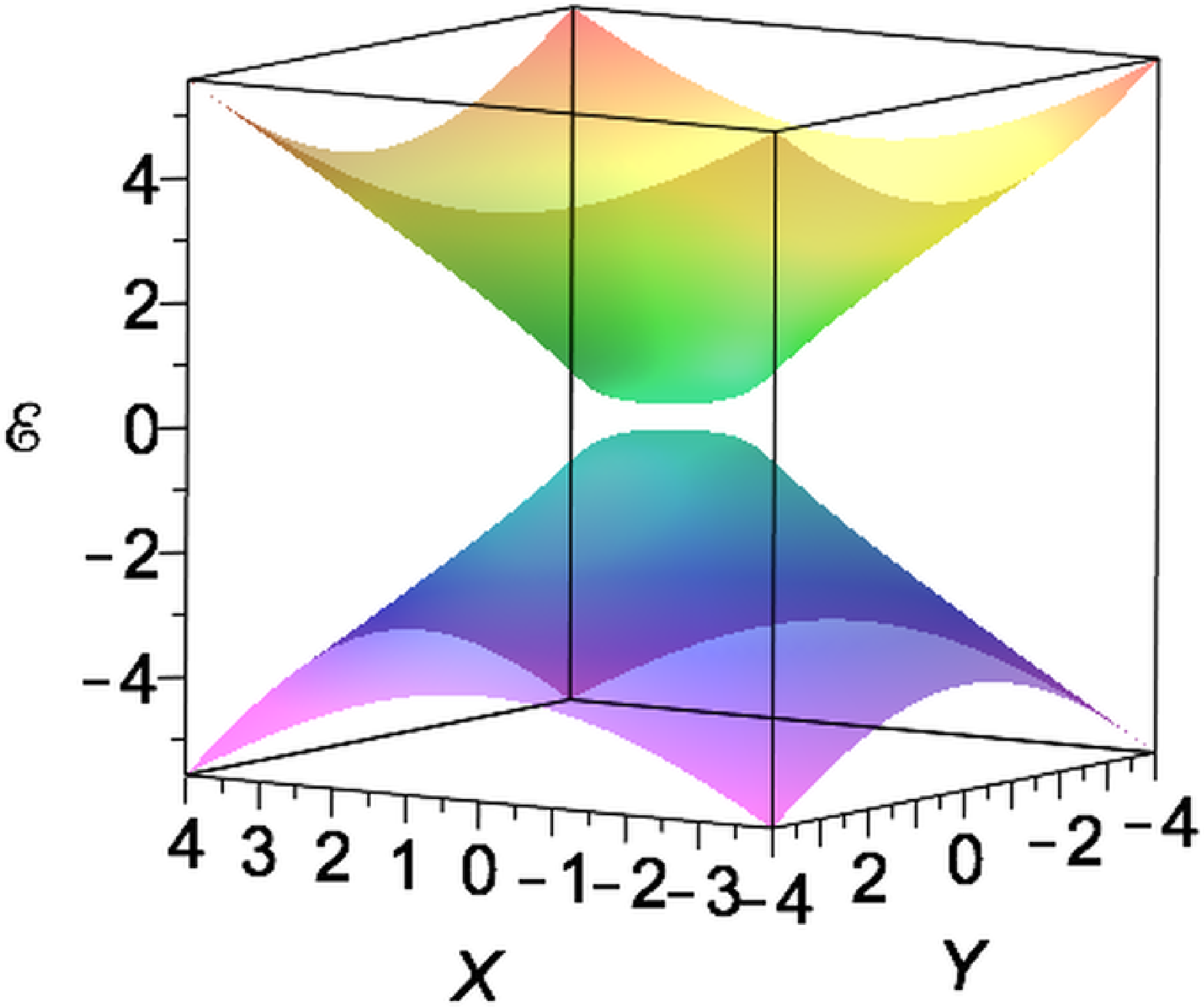}}	
\caption{(Color online) The behaviors of the real parts of the eigenvalues, ${\mathcal E}_\alpha= \Re (E_\alpha)$, in the vicinity of the EP,  for $\Gamma =1$. Left panel: $\varepsilon =0$. The  EP is represented by a circle, $X^2  + Y^2= \Gamma^2$. Right panel: $\varepsilon =0.25$.}
\label{EP}
\end{figure}

The effective non-Hermitian operator $ \tilde{\mathcal H}$ has a singularity when 
the eigenvalues $\tilde E_{1}$ and $\tilde E_{2}$ coincide. To describe the behavior of the eigenvalues in the vicinity of the degeneracy, it is convenient to consider the parameters: $X= \Re V$, $Y= \Im V$ and $Z= \varepsilon - i\Gamma$. In the complex parameter space $(X,Y,Z)$, crossing points, known also as {\em exceptional points}  (EP) \cite{KT,H0,H1,B,IR}, are determined by equation, $X^2 +Y^2 + Z^2=0$. In Fig. \ref{EP}, the real parts of the complex eigenenergies, ${\mathcal E}_\alpha$, are depicted. For  $\varepsilon =0$ (left panel),  the eigenvalues coalesce at the EP, represented by the circle, $X^2  + Y^2= \Gamma^2$.  For  $\varepsilon=0.25$ (right panel), the gap between the valence and conduction bands occurs. 

Note, that in contrast to the case of the Hermitian Hamiltonian, where the degeneracy is referred to a ``conical intersection" at the DP and the coalescence of eigenvalues results in different eigenvectors, at the EP the eigenvectors merge, forming a Jordan block. This leads to the violation of the normalization condition at the EP: $\langle \tilde u_{1,2}|u_{1,2}\rangle = 0$.

Let us assume that the renormalization energies are the same for both sublatticies,  $\varepsilon_1 = \varepsilon_2$.  Then, for $|\Gamma| < |V|$, we obtain for the spacing, $\Delta{\mathcal E} = \sqrt{X^2+ Y^2 - \Gamma^2} $, and  both eigenstates have the same widths. From here, it follows that the real  parts of the eigenenergies, ${\mathcal E}_1 $ and ${\mathcal E}_2$, attract each other as $|\Gamma|$ grows.  At the EP the levels cross, and the complex eigenenergies loose their analytical properties \cite{VZA,VZ1,VZ2,Zel1,VZ3}. 

For $|\Gamma| \geq |V|$, the real parts of eigenenergies become equal,  and the decay widths of the resonances  separate, $|\Upsilon_1- \Upsilon_2| = \sqrt{\Gamma^2- X^2- Y^2 } $ (see Fig. \ref{D1a}, solid lines).  In particular, for $|\Gamma| \gg |V|$,  we obtain
$\Upsilon_{1} = {\Gamma_1}/{2} + {\mathcal O}((|V|/|\Gamma|)^2)$ and $\Upsilon_{2} = {\Gamma_2}/{2} +  {\mathcal O}((|V|/|\Gamma|)^2)$.

 Thus, if $\Gamma_1\gg \Gamma_2$, one has a superradiant eigenstate with $\tilde E_1$, and the subradient long-living eigenstate with  $\tilde E_2$. In the opposite case, $\Gamma_2\gg \Gamma_1$, the eigenstate with $\tilde E_1$ becomes subradient long-living,  and the eigenstate with $\tilde E_2$ becomes a superradiant rapidly decaying eigenstate.
     
\section{The electron transfer into sinks}

The ET quantum dynamics can be described by the Liouville-von Neumann equation,  
\begin{eqnarray}\label{DM1}
i\dot{ \rho} = i[\rho,\mathcal H] - \{\mathcal W,\rho\},
\end{eqnarray}
where $\{\mathcal W,\rho\}= \mathcal W\rho +\rho\mathcal  W$. We define the ET efficiency of tunneling into $n$-th sink as \cite{Lloyd1,CDCH},
\begin{eqnarray}
\eta_n(t) = \Gamma_n \int_0^t \rho_{nn}(\tau)d \tau, \quad n=1,2.
\label{Eq16ar}
\end{eqnarray}
Employing Eq. (\ref{DM1}), one can show that  the following normalization condition is satisfied, $ {\rm Tr}\rho(t) +\eta(t) =1$. Here, $\eta(t) =\int_0^t {\rm Tr}\{\mathcal W,\rho(\tau)\} d \tau =  \eta_1(t)+\eta_2(t)$, is the total ET efficiency of tunneling into both sinks.

\begin{figure}[ht]
 \scalebox{0.35}{\includegraphics{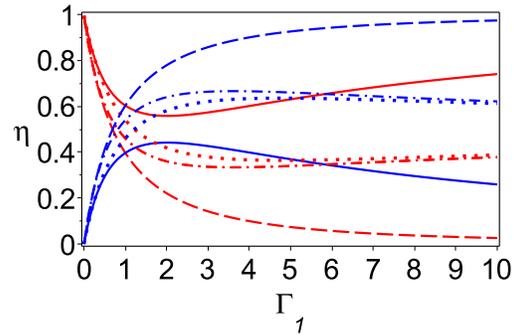}}	
 \caption{(Color online) The efficiencies of sinks, $\eta_1$ (blue) and $\eta_2$ (red)  vs ${\Gamma_1}$, in the absence of noise. Initial conditions: $\rho_{11}(0) =0$,  $\rho_{12}(0) =\rho_{21}(0) =0$  $\rho_{22}(0) =1$ (solid curves); $\rho_{11}(0) =1$,  $\rho_{12}(0) =\rho_{21}(0) =0$  $\rho_{22}(0) =0$ (dashed curves); $\rho_{11}(0) =0.5$,  $\rho_{12}(0) =\rho_{21}(0) =0.5$  $\rho_{22}(0) =0.5$ (dot-dashed curves); $\rho_{11}(0) =0.5$,  $\rho_{12}(0) =\rho_{21}(0) = -0.5$  $\rho_{22}(0) =0.5$ (dotted curves). Parameters: $\Gamma_2 =1$, $X =2$, $Y =0$, $\varepsilon =0.1$. Time of evolution: $\tau = 10 \, \rm ps$.
 \label{C1}}
 \end{figure}

 In Fig. \ref{C1}, the efficiencies, $\eta_1$ (blue curves) and $\eta_2$ (red curves), are shown as functions of $\Gamma_1$,   for fixed values of $\Gamma_2=1$ and  $\varepsilon=0.1$. The time of evolution was chosen, $\tau = 10 \, \rm ps$. The choice of the time-interval, $\tau$, was based on the numerical experiments  which demonstrated that during this time the efficiencies of sinks reached their asymptotic values.
 
If the sublattice B is initially populated, the efficiency $\eta_1$ (blue solid line)  experiences a maximum at $\Gamma_1\approx 2$. This maximum is associated with the superradiance into first sink connected with the sublattice A, and it can be qualitatively explained as follows. The initial population of the sublattice B corresponds to the site state, $|2\rangle$. When $\Gamma_1\ll\Gamma_2$, the ET mainly occurs into the second sink. So, $\eta_2$ is close to 1 and $\eta_1\ll 1$ in this region. When $\Gamma_1$ increases, the ET occurs with maximum in $\eta_2$ (blue curve) and minimum in $\eta_1$ (red curve). When $\Gamma_1$ further  increases, the superradiant eigenstate (with large decay width, $\Gamma_1$) is mainly localized on the site state,  $|1\rangle$, which only weakly overlaps with the initially populated state $|2\rangle$. So, the efficiency, $\eta_1$, decreases for large values of $\Gamma_1$. 
  
  As Fig. \ref{C1} shows, even if only the conduction band (dot-dashed curves)   or the valence band (dotted  curves) of the Hamiltonian (\ref{GH}), is initially populated, the maximum in $\eta_1$ can still be observed. Evidently, when the site state, $|1\rangle$, is initially populated, the maximum in $\eta_1$ does not occur. 

\begin{figure}[ht]
\scalebox{0.165}{\includegraphics{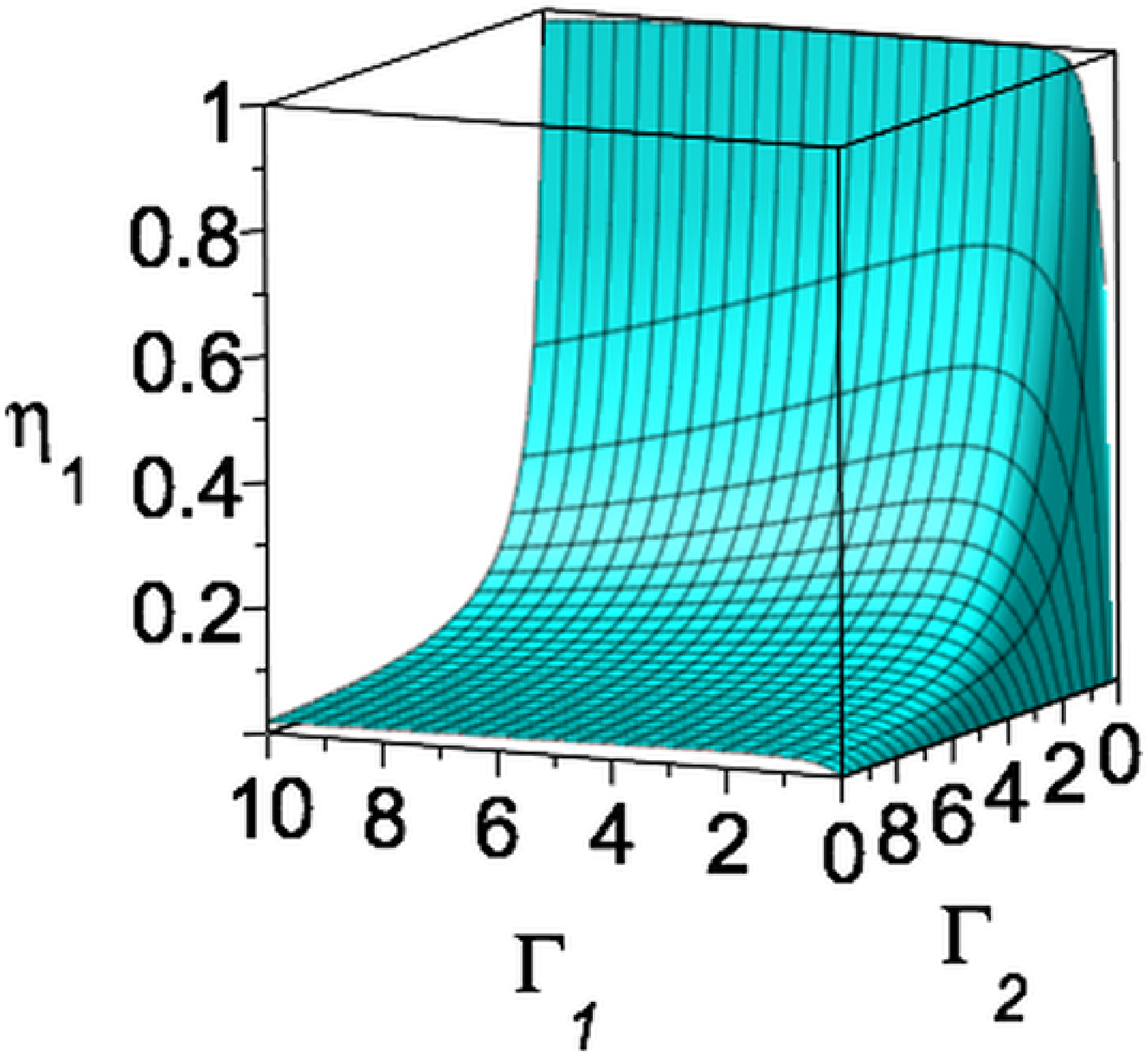}}	
\scalebox{0.165}{\includegraphics{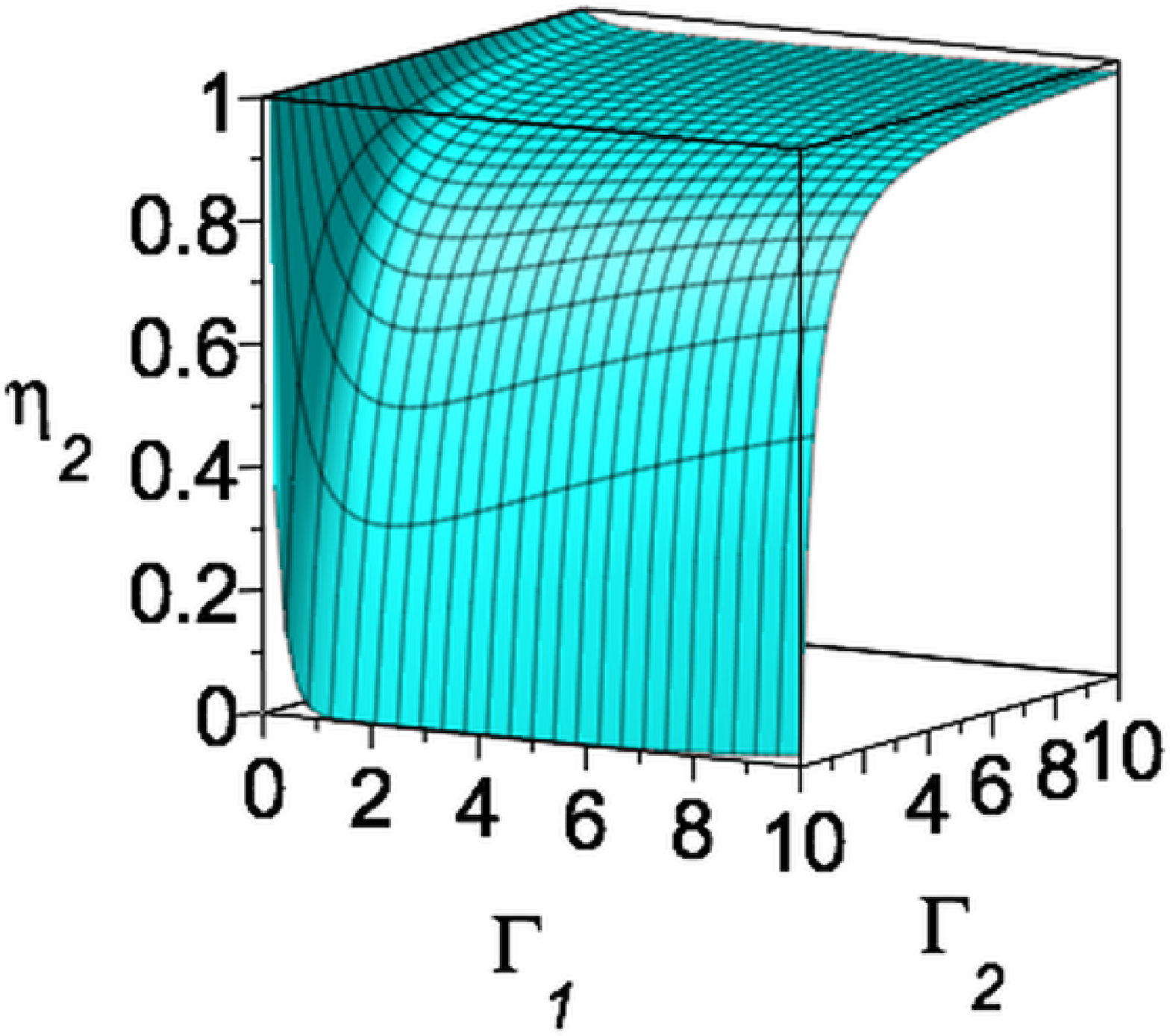}}	
 \caption{(Color online) The efficiencies of the sinks  vs ${\Gamma_1}$ and ${\Gamma_2}$, in the absence of noise.  Initial conditions: $\rho_{11}(0) =\rho_{12}(0) =\rho_{21}(0) =0$,  $\rho_{22}(0) =1$. Parameters:  $|V|=2$, $\varepsilon =0.1$. Time of evolution is: $\tau = 10 \, \rm ps$.
 \label{A2}}
 \end{figure}

In Fig. \ref{A2}, the efficiencies, $\eta_1$ and $\eta_2$, are demonstrated as functions of the ET rates, $\Gamma_1$ and $\Gamma_2$, for initially population of the sublattice B, and for relatively small $\varepsilon=0.1\rm ps^{-1}\approx 0.066\rm meV$.
 
Using the results obtained in \cite{NB4}, we derive the analytical expressions for the efficiencies, $\eta_1$ and $\eta_2$. For illustrative purposes, in what follows, we restrict ourselves by some simple cases. Let us  assume that initially the sublattice B is populated. Then, the  efficiency of trapping the electron into the  first sink  is given by, 
\begin{eqnarray}
& \eta_1(t) = \eta_0-  e^{-\Gamma_0 t}\big(B (\Gamma_0\cosh{\Omega_2 t} 
 + \Omega_2\sinh{\Omega_2 t}) \nonumber \\
& - C (\Gamma_0\cos{\Omega_1 t} -\Omega_1\sin{\Omega_1 t} )\big),
\label{eta}
\end{eqnarray}
where, $\Omega_1 = \Re \Omega$, $\Omega_2 = \Im \Omega$, and 
\begin{eqnarray}
\eta_0=\frac{|V|^2\Gamma_0\Gamma_1}{2(\Gamma^2_0 + \Omega^2_1)(\Gamma^2_0 - \Omega^2_2)}, \\
B=\frac{|V|^2\Gamma_1}{2(\Omega^2_1 + \Omega^2_2)(\Gamma^2_0 - \Omega^2_2)}, \\
C=\frac{|V|^2\Gamma_1}{2(\Omega^2_1 + \Omega^2_2)(\Gamma^2_0 + \Omega^2_1)}.
\end{eqnarray}
\begin{figure}[ht]
\scalebox{0.185}{\includegraphics{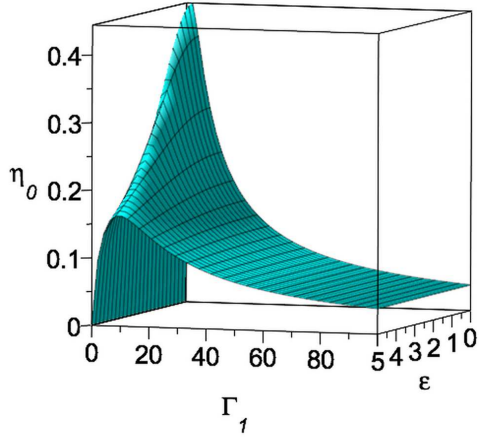}}	
\caption{(Color online) The asymptotic efficiency of the 1st sink, $\eta_1(\infty)=\eta_0$,  vs ${\Gamma_1}$ and ${\varepsilon}$, in the absence of noise.  Initial conditions: $\rho_{11}(0) =\rho_{12}(0) =\rho_{21}(0) =0$,  $\rho_{22}(0) =1$. Parameters:  $|V|=2$, $\Gamma_2 =1$. 
\label{A7}}
\end{figure} 
 \begin{figure}[ht]
 \scalebox{0.35}{\includegraphics{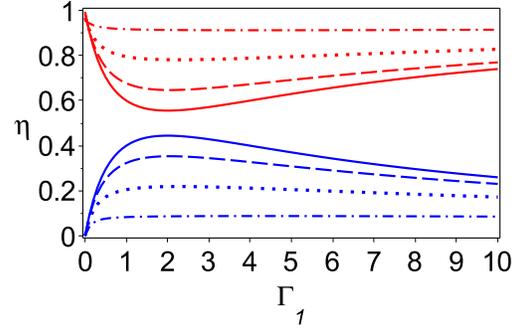}}	
 	\caption{(Color online) The efficiencies of sinks with noise, $\eta_1$ (blue) and $\eta_2$ (red)  vs ${\Gamma_1}$: $d=0$ (solid curves), $d=5$ (dashed curves), $d=10$ (dotted lines), $d=20$ (dot-dashed curves). Parameters: $\gamma=10$, $\Gamma_2 =1$, $|V|=2$,  $d=d_1-d_2$, $\varepsilon =0$.  Initial conditions: $\rho_{11}(0) =0$,  $\rho_{12}(0) =\rho_{21}(0) =0$  $\rho_{22}(0) =1$. Time of evolution: $\tau = 10 \, \rm ps$.
 		\label{A6}}
 \end{figure}
 
For $|\Gamma t |\gg 1$, the asymptotic behaviors of $\eta_{1,2}$ are given by, 
\begin{eqnarray}\label{Eq17a}
\eta_1(t) \approx \eta_0-D e^{-2\Upsilon_2 t},\\
\eta_2(t) \approx 1-\eta_0+D e^{-2\Upsilon_2 t},
\end{eqnarray}
where,
\begin{eqnarray}
D= \frac{|V|^2\Gamma_1}{4(\Omega^2_1 + \Omega^2_2)(\Gamma_0 - \Omega_2)}.
\end{eqnarray}
For fixed $\Gamma_2$, the  behavior of the asymptotic value, $\eta_0$, as the function of $\Gamma_1$ and $\varepsilon$ is demonstrated in Fig. \ref{A7}. One can see the details of the maximum of the efficiency, $\eta_0$, in the first sink depending on both parameters. The most pronounced maximum of $\eta_0$ corresponds to the graphene case -- absence of the effective electron mass, $\varepsilon=0$.

\section{Noise-assisted electron transfer}

In the presence of noisy environment, the evolution of the system can be described by the following effective non-Hermitian Hamiltonian \cite{NB1,NB2,BNSS}, 
\begin{eqnarray}
\tilde{\mathcal H}_{tot}= {\mathcal H}- i \mathcal W + {\mathcal V}(t),
\end{eqnarray}
where, ${\mathcal V}(t)=  \sum_{m,n} \lambda_{mn}(t)|m\rangle\langle  n |, \quad m,n = 1,2$, and $\lambda_{mn}(t)$ describes the influence of the noisy environment. In what follows, we restrict ourselves with the diagonal noise,  writing $\lambda_{mn}(t) = \lambda_n \delta_{mn}\xi(t) $, where $\lambda_n$ is the coupling constant at site, $n$, and $\xi(t)$ is the random telegraph process (RTP) with the properties \cite{NB1,NB2,BNSS}, 
\begin{eqnarray}
\langle \xi(t)\rangle =0, \quad \chi(t-t')=\langle \xi(t)\xi(t')\rangle, 
\end{eqnarray}
where, $\chi(t-t') = \sigma^2 e^{-2\gamma |t-t'|}$, is the correlation function of noise. 

The evolution of the  average  components of the density matrix is described by the following closed system of ordinary differential equations \cite{NB2,BNSS}:
\begin{align} \label{IB4}
\frac{d}{dt}{\langle{\rho}}\rangle =&i[\langle\rho\rangle,\mathcal H] - \{\mathcal W,\langle\rho\rangle\} - i B\langle\rho^\xi\rangle, \\
\frac{d}{dt}{\langle{\rho^\xi}}\rangle =&i[\langle\rho^\xi\rangle,\mathcal H] - \{\mathcal W,\langle\rho^\xi\rangle  \nonumber \\
&- i B\langle\rho\rangle- 2\gamma \langle\rho^\xi\rangle ,
\label{IB5}
\end{align}
where $\langle\rho^\xi\rangle = \langle\xi\rho \rangle/\sigma$, $B = \sum_{m,n}(d_m - d_n)|m\rangle \langle n|$, and we set $d_{m}=\lambda_m \sigma$. The average,  $\langle \,\dots \,\rangle$, is taken over the RTP. 

In Fig. \ref{A6}, the influence of noise on the ET rate is demonstrated, for the effective electron mass,  $\varepsilon=0$, and for fixed parameters: $\Gamma_2=1$,  $\gamma=10$ and $|V|=2$. The amplitude of noise is characterized by the parameter, $d=d_1-d_2$. As one can see from Fig. \ref{A6}, the ET is stable in the sense that when $d\neq 0$, the maximum in $\eta_1$ still exist, but is less pronounced, when $d$ increases.  

\section{Conclusion}

We studied theoretically and numerically a possibility to observe a superradiance transition in a  single-layer graphene material. The analysis was performed in the tight-binding approximation and at the vicinity of the zero gap between the conduction and the valence bands. For each triangle sublattice  we introduced the electron transfer rates  in their continuum (sinks). So, the total model, in the ${\mathbf k}$-representation,  is reduced to the two-level system governed by the non-Hermitian Hamiltonian.

In the graphene case, the superradiance transition occurs in the vicinity of the exceptional point, where two complex eigenenergies coincide. Relatively far from the exceptional point, a qualitative interpretation of the superradiance transition is associated with the overlapping of two resonances: when the sum of half-widths (decay rates) of resonances is of the order of the spacing between them (distance between the real parts of the complex eigenenergies). When one of the electron transfer rates further increases, a segregation of the resonances  occurs. Namely, one decay rate, corresponding to the superradiant eigenstate, grows and the second one, corresponding to the subradient eigenstate,  becomes limited from above. 

The superradiant state is usually responsible for rapid coherent electron transfer into a continuum (sink). Under some conditions on parameters and initial population, which are analyzed in the paper, the efficiency of the ET into a sink has a maximum at the  particular value of the electron transfer rate.

In order to observe the described superradiance  in the graphene material, some additional experimental issues must be  resolved. In particular, it is required to realize and control two electron transfer rates for two sublattices. It is also required to prepare the system in the particular initial states, say, to populate initially one of the sublattices. The experimental observation of the discussed superradiance transition will be important for both, better understanding of the fundamental properties  and for controlling of the graphene based material.

\begin{acknowledgements}  	
   A.I.N.  acknowledges the support from the CONACyT. 
\end{acknowledgements}

\end{document}